\documentclass[final,5p,times,twocolumn,authoryear]{elsarticle}

\usepackage{amssymb}
\usepackage{lscape}
\usepackage{truncate}

\usepackage{natbib}
\usepackage{longtable}
\bibpunct{(}{)}{;}{a}{,}{,}
\usepackage{aas_macros}

\usepackage{graphicx}
\usepackage{pifont}
\usepackage{ulem}
\usepackage{bbding}

\newlength{\titleLength}
\newlength{\titleOffset}

\usepackage{pstricks}
\usepackage{color}
\graphicspath{ {figures/} }

\usepackage{hyperref}
\hypersetup{
    bookmarks=true,         % show bookmarks bar?
    unicode=true,           % non-Latin characters in Acrobat's bookmarks
    pdftoolbar=true,        % show Acrobat's toolbar?
    pdfmenubar=true,        % show Acrobat's menu?
    pdffitwindow=true,      % page fit to window when opened
    pdftitle={ViSiON: Visibility Service for Observing Nights},
    pdfauthor={B. Carry and J. Berthier},
    pdfsubject={Planetary Science},
    pdfkeywords={},         % list of keywords
    pdfnewwindow=true,      % links in new window
    colorlinks=true,        % false: boxed links; true: colored links
    linkcolor=gray,         % color of internal links
    citecolor=blue,         % color of links to bibliography
    filecolor=gray,         % color of file links
    urlcolor=gray           % color of external links
}

\newcommand{\numb}[1]{\textcolor{orange}{#1}}
\newcommand{\degr}{\ensuremath{^{\circ}}}
\newcommand{\arcsec}{\ensuremath{^{\prime\prime}}}

\renewcommand{\numb}[1]{#1}

\newcommand{\vision}{\texttt{ViSiON}}
\newcommand{\skybot}{\texttt{Skybot}}
\newcommand{\miriade}{\texttt{Miriade}}
\newcommand{\simbad}{\texttt{SIMBAD}}

\newcommand{\tsup}[1]{\ensuremath{^{\textrm{\small #1}}}}
\newcommand{\tsub}[1]{\ensuremath{_{\texttt{\small #1}}}}

\newcommand{\rem}[1]{}
\newcommand{\add}[1]{\textbf{#1}}

\renewcommand{\add}[1]{#1}

\journal{PSS}

\begin{document}

\begin{frontmatter}

%% Title, authors and addresses

%% use the tnoteref command within \title for footnotes;
%% use the tnotetext command for theassociated footnote;
%% use the fnref command within \author or \address for footnotes;
%% use the fntext command for theassociated footnote;
%% use the corref command within \author for corresponding author footnotes;
%% use the cortext command for theassociated footnote;
%% use the ead command for the email address,
%% and the form \ead[url] for the home page:
%% \title{Title\tnoteref{label1}}
%% \tnotetext[label1]{}
%% \author{Name\corref{cor1}\fnref{label2}}
%% \ead{email address}
%% \ead[url]{home page}
%% \fntext[label2]{}
%% \cortext[cor1]{}
%% \address{Address\fnref{label3}}
%% \fntext[label3]{}

\title{\vision: Visibility Service for Observing Nights}

%% use optional labels to link authors explicitly to addresses:
%% \author[label1,label2]{}
%% \address[label1]{}
%% \address[label2]{}

\author[oca,imcce]{B. Carry}
\author[imcce]{J. Berthier}

\address[oca]{Universit\'e C{\^o}te d'Azur, Observatoire de la
  C{\^o}te d'Azur, CNRS, Laboratoire Lagrange, France}
\address[imcce]{IMCCE, Observatoire de Paris, PSL Research University, CNRS,
  Sorbonne Universit{\'e}s, UPMC Univ Paris 06, Univ. Lille, France}

%--------------------------------------------------------------------------------------------------
%--------------------------------------------------------------------------------------------------
%--------------------------------------------------------------------------------------------------
%--------------------------------------------------------------------------------------------------
\begin{abstract}
  Preparation of detailed night schedule prior to an observing run
  can be tedious, especially for solar system objects which
  coordinates are epoch-dependent.
  We aim at providing the community with a Web service
  compliant with Virtual Observatory (VO) standards, to create tables of
  observing conditions, together with airmass and sky charts, for an
  arbitrary list of targets.
  We take advantage of available VO services such as 
  the \simbad~astronomical database,
  the Aladin sky atlas, and
  the \miriade~ephemerides generator to build a new service
  dedicated to the planning of observations.
  The requests for ephemerides charts and tables are handled by a
  VO-compliant Web service. For each date, and each target,
  coordinates in local and equatorial frames are computed, and used
  to select targets accordingly to user's criteria for visibility.
  This new service, dubbed \vision~for 
  \textbf{Vi}sibility
  \textbf{S}erv\textbf{i}ce for
  \textbf{O}bserving
  \textbf{N}ight, is a new method of \miriade~Web service hosted at
  IMCCE.
  It allows anyone to create graphics of observing conditions and tables 
  summarizing them, provided as PDF, VOTable, and xHTML documents.
\end{abstract}

\begin{keyword}
%% keywords here, in the form: keyword \sep keyword
ephemerides \sep
virtual observatory tools \sep
methods: numerical 

%% PACS codes here, in the form: \PACS code \sep code

%% MSC codes here, in the form: \MSC code \sep code
%% or \MSC[2008] code \sep code (2000 is the default)

\end{keyword}

\end{frontmatter}

% \linenumbers

%\tableofcontents

%%------------------------------------------------------------------------------------------
%%------------------------------------------------------------------------------------------
%%------------------------------------------------------------------------------------------
%%------------------------------------------------------------------------------------------
%%------------------------------------------------------------------------------------------
\section{Introduction}
  \indent To prepare and conduct observations
  from a ground-based telescope, many astronomers create airmass
  charts, also called visibility plots. These graphics represent 
  the elevation above the horizon, or altitude, 
  of each target during the night, providing a graphical and easy way
  to schedule the sequence of observations, in which each target
  should be observed as closely as possible to its highest altitude.
  Another type of visual guidelines are fish-eye sky charts, in which
  the successive positions of the targets are displayed by tracks on a
  stereographic projection of the local frame. It also helps
  optimizing observations, by avoiding pointing the telescope back and
  forth from different regions of the sky. \\
  \indent If many astronomers have developed their own piece of code
  for that purpose, several efforts have been made to make such tools freely
  available to the community. The \texttt{staralt}
  pages\footnote{\href{http://catserver.ing.iac.es/staralt/}
    {http://catserver.ing.iac.es/staralt/}}, by J. M{\'e}ndez, 
  allows productions of such graphics.
  The \texttt{skycalc}
  software\footnote{\href{http://www.dartmouth.edu/~physics/labs/skycalc/flyer.html}
      {http://www.dartmouth.edu/~physics/labs/skycalc/flyer.html}}, maintained
  by J. Thorstensen, represents an easily-portable and interactive
  solution to these points. Yet, these tools have been developed for objects 
  with fixed coordinates, and the eight planets. Planning observations of moving 
  Solar System Objects (SSOs) may therefore result tedious: one should
  compute first the coordinates of each object, and then enter them into
  the software interface.
  This often results impracticable if the list
  of targets is long, or many dates are envisioned. \\
  \indent We aim here at providing the community with a simple Web service,
  fully compliant with Virtual Observatory (VO) standards, 
  to produce tables and charts of visibility for both fixed
  coordinates targets and moving solar system objects.
  This
  service\footnote{\add{\url{http://vo.imcce.fr/webservices/miriade/?vision}}}
    is dubbed
  \textbf{Vi}sibility
  \textbf{S}erv\textbf{i}ce for
  \textbf{O}bserving
  \textbf{N}ight
  (\vision).
  Targeted users are observers willing to prepare an upcoming night,
  or to check the observability of some targets on-the-fly, and
  developers of night scheduler and telescope control software willing
  to include airmass charts in graphical interfaces. \\
  \indent The present article is structured as following:
  we quickly describe the core of ephemeris computation in
  Sect.~\ref{sec:ephem}.
  The algorithm of \vision~is detailed in Sect.~\ref{sec:vision}, and
  the criteria which define observability are presented in 
  Sect.~\ref{sec:select}. The different ways to access to the
  service are provided in Sect.~\ref{sec:io} and its links with other
  VO services in Sect.~\ref{sec:links}

\section{Ephemeris computation}
\label{sec:ephem}
  \rem{The \textsl{Bureau des longitudes},
    created during the French revolution by the law of the
    Messidor 7, year 3 by the Convention Nationale, is the academy responsible of the
    definition of the French national ephemerides. The practical
    realization of these ephemerides is entrusted to the
    \textsl{Institut de
      m\'ecanique c\'eleste et de calcul des \'eph\'em\'erides} 
    (IMCCE), affiliated with the Paris Observatory. \\}
  \indent The ephemerides of planets and small Solar system objects
  are computed into a quasi-inertial reference frame, taking into
  account the post-Newtonian approximations.
  The geometric positions of the major planets and the Moon are
  provided by INPOP planetary theory \citep{2014-SciNote-Fienga}.
  Those of small SSOs (asteroids, comets, Centaurs, trans-neptunian
  objects) are calculated by 
  numerical integration of the N-body perturbed problem
  \citep[Gragg-Bulirsch-Stoer algorithm, see][]{1966-Bulirsch,
   1980-Bulirsch}, using their latest published 
  osculating elements, from the 
  \texttt{astorb} \citep{1993-LPI-Bowell} and
  \texttt{cometpro} \citep{1996-IAUS-Rocher}  
  databases. %
  The overall accuracy of asteroid and comet ephemerides provided by \vision~is 
  at the level of tens of milli-arcseconds, mainly depending on the accuracy of
  their osculating elements. The positions of natural satellites
  are obtained thanks to dedicated solutions of their motion, 
  e.g., \citet{2004-AA-427-Lainey, 2004-AA-420-Lainey, 2007-AA-465-Lainey} for
  Mars and Jupiter, \citet{1995-AA-297-Vienne} for Saturn,
  \citet{1987-AA-188-Laskar} for Uranus, and 
  \cite{1993-AA-272-LeGuyader} for Neptune satellites. \\
  \indent The ephemerides of stars, and more generally of all
  \simbad~objects 
  \add{\citep{2000-AA-143-Wenger}}, 
  are also computed into a quasi-inertial reference frame, taking into account 
  their proper motion, parallax and radial velocity, as long as the corresponding 
  input parameters are provided by the catalogue. The ephemeris accuracy of 
  \simbad~objects is thus at the level of the accuracy of their catalogued position. \\
  \indent All the ephemerides in \vision~are computed relatively to a given observer,
  in the ICRF reference frame (mean J2000 equator and equinox), and expressed in the 
  UTC time scale. The coordinates of the observer (the topocenter) can be either provided 
  directly by users (longitude, latitude, altitude), or by using the observatory code 
  provided by IAU Minor Planet
  Center\footnote{\href{http://www.minorplanetcenter.net/iau/lists/ObsCodes.html}{http://www.minorplanetcenter.net/iau/lists/ObsCodes.html}} 
  (MPC)
  for listed observatories.

\section{Description of \vision's algorithm\label{sec:algo}}
\label{sec:vision}

\begin{figure*}[t]
    \centering
    \includegraphics[width=\hsize]{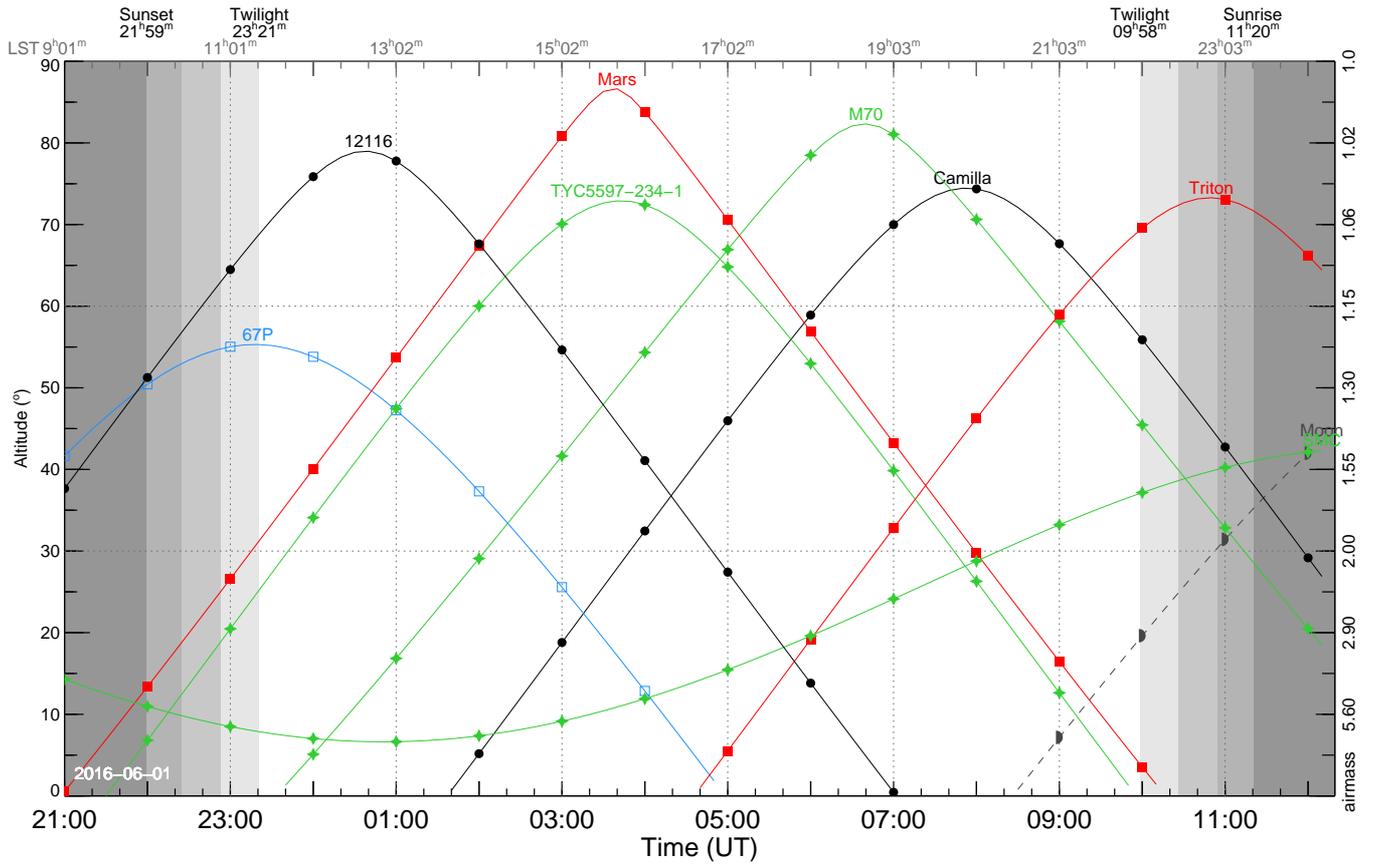}
    \caption{Example of airmass chart produced by \vision~for a
      request a La Silla observatory (ESO) on June the
      1\tsup{st}, 2016.
      The different levels of gray shading on each side correspond to
      the day light, civil, nautical, and astronomical twilights.
      Both universal time (UT) and local sidereal time (LST) are reported
      on x-axes, and altitude and airmass are provided on y-axes.
      Each curve represents a target, color-coded as a function of
      type: \add{
        planet (Mars),
        asteroid (12\,116 and Camilla), 
        comet (67P), 
        satellite (Triton), and
        fixed-coordinates such as
        galaxy (Small Magellanic Cloud: SMC), 
        star (TYC5597-234-1), or
        clusters (M70).
        }
      \label{fig:airmass}
    }
  \end{figure*}

\begin{figure*}[t]
    \centering
    \includegraphics[width=\hsize]{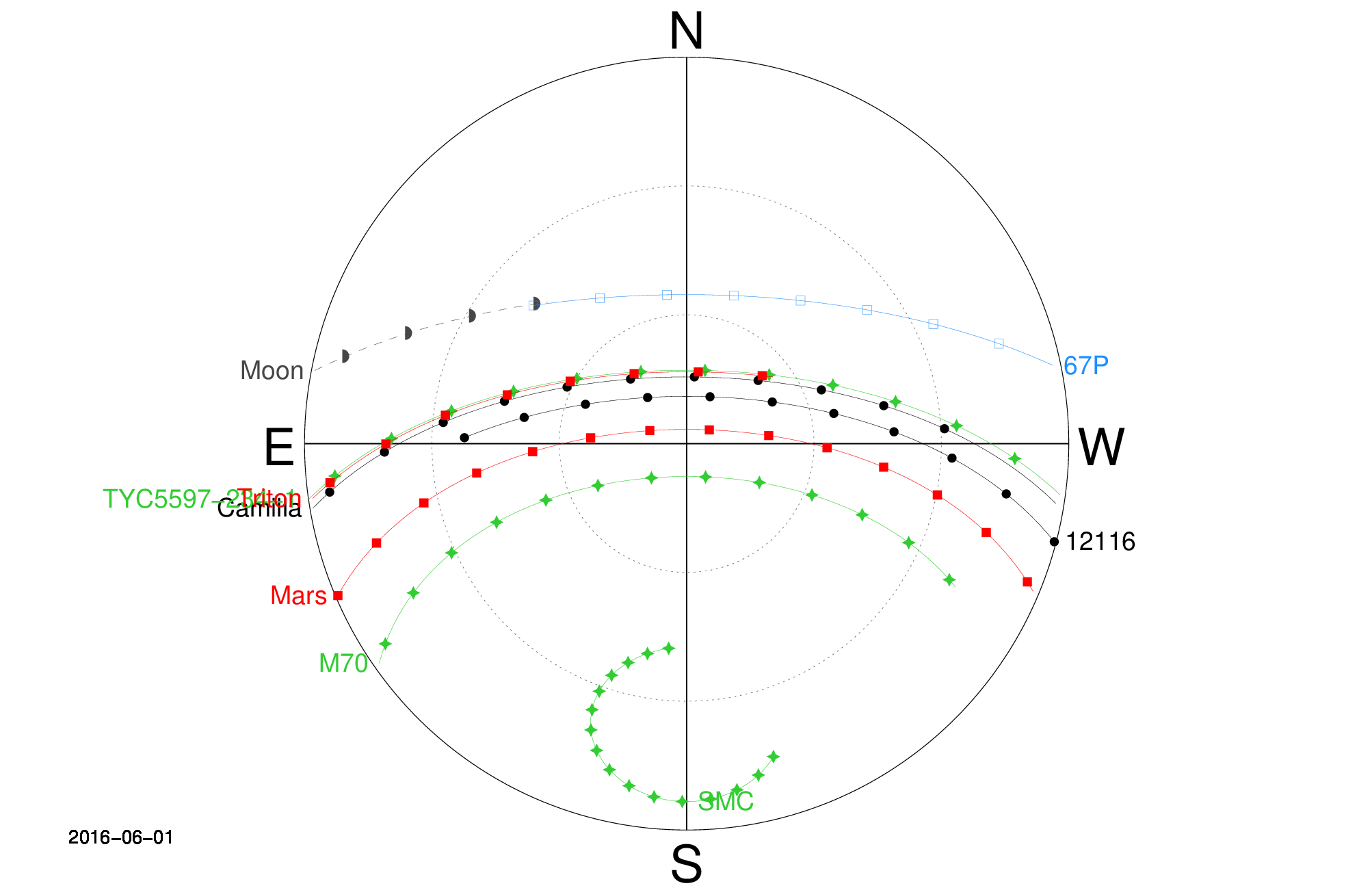}
    \caption{Example of sky chart produced by \vision~for a
      request a La Silla observatory (ESO) on June the
      1\tsup{st}, 2016.
      As in Fig.~\ref{fig:airmass}, each curve represents a target,
      color-coded as a function of its type.
      \label{fig:sky}
    }
  \end{figure*}

  \indent For each requested date, a series of computations are
  performed prior to selection of targets suitable for observation.
  First, the ephemerides of the Sun are computed, to determine the
  \add{instants of sunset and sunrise}. The limits of
  the civil, nautical, and astronomical twilights, corresponding to the times
  when the altitude of the Sun is 6\degr, 12\degr, and 18\degr~below the horizon,  are
  also computed at this stage.\\
  \indent Second, the phase and position of the Moon \add{throughout} the night 
  are computed and stored. Finally, the topocentric astrometric coordinates 
  of each target are computed, with a time step of 5 minutes, and expressed 
  as equatorial (RA, Dec), horizontal (Az, Alt), and galactic coordinates 
  ($\lambda_G$, $\beta_G$). \\
  \indent Additionally, both solar and lunar
  elongations are computed at each time step, together with 
  \add{(for Solar system objects only)} the
  solar phase angle, the heliocentric distance and the range to the
  observer.
  These detailed coordinates are used to determine for how long each
  target is visible during the night. The \add{entire} list of criteria
  defining observability is provided in Sec.~\ref{sec:select}.
  If there is at least one target observable, the night is included in
  the result, and these detailed
  coordinates are the inputs to the airmass and
  sky charts (see Figs.~\ref{fig:airmass} and~\ref{fig:sky} for
  examples). 
  For each date, a table lists all the targets that are visible
  \add{(see Table~\ref{tab:ex} for an example)}, and
  summarizes their observing conditions at the time of highest
  altitude \add{(see Table~\ref{tab:col} for a description of each
    parameter)}.

%-------------------------------------------------------------
\begin{table}[t]
  \caption{List of parameters reported for each visible target, for
    each requested night. The values reported in \vision~correspond to
    the geometry at the time of the highest altitude on sky.
    \label{tab:col}
  }
  \centering
    \begin{tabular}{lll}
      \hline\hline
      Quantity & Description & Units\\
      \hline
      Target              & Target designation                 & --       \\
      m$_V$               & Apparent magnitude                 & mag      \\
      $\phi$              & Apparent diameter                  & arcsec   \\
      $\mathcal{D}$       & Duration of visibility window      & h:m      \\
      Alt                 & Altitude                           & deg      \\
      Az.                 & Azimuth                            & deg      \\
      RA                  & Right Ascension                    & h:m:s    \\
      DEC                 & Declination                        & d:m:s    \\
      Rate                & Apparent non-sidereal motion       & arcsec/h \\
      $\lambda_G$         & Galactic longitude                 & deg      \\
      $\beta_G$           & Galactic latitude                  & deg      \\
      $r$                 & Range to observer                  & au       \\
      $\Delta$            & Heliocentric distance              & au       \\
      $\alpha$            & Solar phase angle                  & deg      \\
      $\widehat{\rm SEO}$ & Solar elongation                   & deg      \\
      $\widehat{\rm MEO}$ & Moon elongation                    & deg      \\
      \hline
    \end{tabular}
\end{table}
%-------------------------------------------------------------

%-------------------------------------------------------------
%-------------------------------------------------------------

%-------------------------------------------------------------
%-------------------------------------------------------------

\section{Selection and sorting criteria}
\label{sec:select}

  \indent A suite of parameters are tested to determine if a target
  shall be considered as visible. The thresholds for selection are
  provided as inputs, accordingly to each observer's preference.
%  (using the \texttt{cuts} parameter, see~Table~\ref{tab:param}).
  The following parameters can be specified
  (see Table~\ref{tab:cuts} for details):
  altitude, apparent magnitude, solar phase angle, 
  solar elongation, lunar elongation, angular diameter, duration of
  visibility window\add{,
    and events limiting the ``night'' (sunrise/sunset, or
    civil/nautical/astronomical twilights)}. \\
  \indent \vision~proceeds as following: for each target, it computes
  its ephemerides for each observing date, as described above
  (Sect.~\ref{sec:algo}). \add{Each parameter (apparent magnitude, elevation,
  etc) is checked against the selection criteria at every 5\,min time
  step.}
  A time step is considered valid \add{only if the tests against
  all the criteria have passed}. Finally, the target will be considered
  valid for the date if the
  number of time steps exceeds the specified minimum duration of
  visibility (30\,min by default)
  \add{between the two events defining the ``night''}.\\
  \indent For each date, all the targets that pass the selection are
  drawn on the graphics, and included in the summary table.
  Their order can also be chosen by users, among all the
  parameters listed in Table~\ref{tab:col}.
%  (\texttt{sort} parameter, see~Table~\ref{tab:param}).
  A double sorting is even possible, with objects being grouped by
  type first (asteroid, comet, planet, satellite, and
  fixed-coordinate) then by the requested parameter.
%  (syntax: \texttt{sort=type$>$Alt})

%
%-------------------------------------------------------------
\begin{table*}[t]
  \caption{Selection criteria in \vision.
  For each, we indicate if a minimum and/or a maximum value can be
  specified, and which are the default limits.
  The phase angle and apparent diameter selection only applies to
  solar system objects, and 
  is disregarded for the other targets.}
  \label{tab:cuts}
  \centering
  \begin{tabular}{llccl}
    \hline\hline
    Quantity & Description & Minimum & Maximum & Defaults \\
    \hline
    Alt.                & Highest altitude reached      & \Checkmark & \Checkmark & [30 to 90\degr] \\
    m$_V$               & Apparent magnitude            & \Checkmark & \Checkmark & [No default] \\
    $\alpha$            & Solar phase angle             & \Checkmark & \Checkmark & [0 to 180\degr]  \\
    $\widehat{\rm SEO}$ & Solar elongation              & \Checkmark &            & [No default] \\
    $\widehat{\rm MEO}$ & Moon elongation               & \Checkmark &            & [No default] \\
    $\phi$              & Apparent diameter             & \Checkmark & \Checkmark & [No default]  \\
    $\mathcal{D}$       & Duration of visibility window & \Checkmark &            & [30 min] \\
    \add{Event}         & \add{Start/End of the ``night''} &         & \Checkmark & \add{[civil]} \\
    \hline
  \end{tabular}
\end{table*}
%-------------------------------------------------------------

\section{Access to the service}
\label{sec:io}

  \indent In 2005, the IMCCE started to implement Virtual Observatory (VO) compliant 
  interfaces to its ephe\-me\-rides services \citep{2005-DPS-Thuillot}. A Web
  portal\footnote{\href{http://vo.imcce.fr/}{http://vo.imcce.fr/}} describe 
  the various services, such as solar system objects identification 
  \citep[\skybot,][]{2006-ASPC-351-Berthier}, or general ephe\-me\-rides computation
  \citep[\miriade,][]{2009-EPSC-Berthier}. All our services are accessible 
  via Web services (based on SOAP and HTTP POST verb) which allow
  anyone to interface her/his own software with the services,
  via HTTP request and Web forms,
  and are integrated in several VO-compliant software such as the well-spread
  Aladin Sky Atlas \citep{2000-AA-143-Bonnarel}. \\
%
%-------------------------------------------------------------
\begin{table*}[t]
  \caption{List of input parameters to \vision~(see Sect.~\ref{sec:io}
    for the list of access points).
    For each, the default (between brackets) and range are specified.
  }
  \label{tab:param}
  \centering
  \begin{tabular}{lll}
    \hline\hline
    Parameter & Description & \add{Range, or example values} \\
    \hline
    \texttt{name}    & Designation of the requested targets & p:Mars, a:Pluto, u:M\_31 \\
    \texttt{observer}& Observer's location (IAU code or geographic coordinates) & [Paris, France]\\
    \texttt{ep}      & Starting epoch &[now] | JD | \add{YYYY-MM-DD} \\
    \texttt{nbd}     & Number of date of computation & [1] $\leq$ \texttt{nbd} $<$ 366 \\
    \texttt{step}    & Time interval (days) between two dates & [1]\\
    \texttt{cuts}    & Criteria for visibility selection & (see Sect.~\ref{sec:select}) \\
    \texttt{sort}    & Parameter used to sort entries & [RA] \\
    \texttt{mime}    & Mime type of the results & votable | html | [pdf] \\
    \texttt{from}    & Name of the caller application &  \\
    \hline
 \end{tabular}
\end{table*}
%-------------------------------------------------------------
%
  \indent There are two ways to use the \vision~web service: by
  writting a client to send requests to the \miriade~server and to
  receive and analyze the response \citep[as for instance the ephemerides
  tool for \add{observations in \textsl{Service Mode} hosted} by ESO, see][]{2016-Messenger-163-Carry},
  or by using a command line 
  interface and a data \add{transfer} program such as \texttt{curl} or
  \texttt{wget}.
  We also propose a simple query form on our SSO VO
  portal\footnote{\add{\href{http://vo.imcce.fr/webservices/miriade/?forms\#Vision}{http://vo.imcce.fr/webservices/miriade/?forms\#Vision}}}.
  In any case, the service accepts nine parameters, described in
  Table~\ref{tab:param}.
  \add{Only the list of targets is mandatory, all the other parameters
  being optional.  We }
  describe them in detail here. 

  \paragraph{Targets (\texttt{name})}
  Planets can be requested by setting the \texttt{name}
  parameter to their name or number,
  preceeded by the ``\texttt{p:}'' prefix, for instance
  \texttt{p:Mars} and \texttt{p:5} to request visibility of Mars and Jupiter.
  Any entry in \texttt{astorb} and \texttt{cometpro} can be
  retrieved, using the ``\texttt{a:}'' and ``\texttt{c:}'' prefixes, e.g., 
  \texttt{a:Ceres}, \texttt{a:21}, or \texttt{c:67P} for the asteroids
  (1) Ceres and (21) Lutetia, and
  the comet 67P/Churyumov-Gerasimenko. 
  All the satellites present in the references listed in
  Sect.~\ref{sec:ephem} are available, though the ``\texttt{p:}''
  or ``\texttt{s:}'' suffixes,
  e.g., \texttt{p:Phobos}, \texttt{s:Enceladus}
  \add{for the corresponding satellites of Mars and Saturn}. \\
  \indent For objects with fixed coordinates, \vision~queries the
  \simbad~service. The targets are therefore
  inherited from \simbad, which currently contains above 8 millions
  entries. These targets are submitted to \vision~using the ``\texttt{u:}''
  prefix, e.g., \texttt{u:Sirius}, \texttt{u:HIP\_87937}, \texttt{u:M\_31}
  \add{for Sirius, the Hipparcos star 87937, and Messier 31}.
  For objects without a catalog entry, it is always possible to
  directly submit equatorial coordinates
  (\add{in decimal hours and degrees in} astrometric J2000) to
  \vision, with the ``\texttt{e:}'' prefix. For instance, the
  equatorial North pole can be queried via \texttt{e:0.0+90.0}.\\
  \indent In all cases, \vision~accepts aliases for each target, that
  can be 
  used to simplify the \add{labels on the} graphics, with the syntax 
  \texttt{prefix:name=alias}.
  \add{For instance
    in Figs.~\ref{fig:airmass} and~\ref{fig:sky}, the Small Magellanic
    Cloud was requested as \texttt{e:0.9-72=SMC}, while Mars was
    requested using \texttt{p:4} and M70 with \texttt{u:M\_70}.
    This option may be useful for minor planets with high numbers and
    provisional designation:
    to avoid the display of 2000 WO137 on the graphics when querying for 
    \texttt{a:123456}, one can force the usage of the IAU number by specifying
    \texttt{a:123456=123456}.}

  \paragraph{Topocenter (\texttt{observer})}
  \add{The location on Earth of the observer can be specified 
    by providing the longitude (0,360\degr~West),
    latitude (-90,+90\degr~North), and altitude (meters above sea level)
    or the IAU MPC\footnote{\href{http://www.minorplanetcenter.net/iau/lists/ObsCodes.html}{http://www.minorplanetcenter.net/iau/lists/ObsCodes.html}}
    observatory code, in which case \vision~retrieves the coordinates
    of the observer from the MPC list.
  For instance, the Paris observatory can be specified as
  \texttt{-2.336537 +48.8364 132.0} or \texttt{007}.}

  \paragraph{Dates of computation (\texttt{ep}, \texttt{nbd},
    \texttt{step})} 
  \add{The \texttt{ep} parameter sets the first (or only) date at which the visibility is
    sought (\texttt{YYYY-MM-DD}).
    If more than one night is desired, their total number can be
    specified with \texttt{nbd} (integer $\geq$1) and the number of skipped
    nights between each by \texttt{step} (integer $\geq$0).
    For instance, four dates, one each week, starting from June, the
    1\tsup{st}, 2016, these parameters should be set to
    \texttt{ep=2016-06-01}, \texttt{nbd=4}, and
    \texttt{step=7}.}

  \paragraph{Selection criteria (\texttt{cuts})}
  \add{Any number of selection criteria can be specified, using the 
    syntax
    \texttt{tag\tsub{1}:value\tsub{1}, ..., tag\tsub{N}:value\tsub{N}}, 
    where the \texttt{tag\tsub{i}} are listed in 
    Table~\ref{tab:cuts}.  
    Tags accepting both a minimum and a maximum value shall be set
    using the following syntax
    \texttt{tag\tsub{i}:\{min:value\tsub{min}\}} and
    \texttt{tag\tsub{i}:\{max:value\tsub{max}\}}
    to set only one extrema,
    or
    \texttt{tag\tsub{i}:\{min:value\tsub{min}, max:value\tsub{max}\}}
    to set both.
    For instance, to limit targets to sources brighter than 
    \numb{V\,=\,20}, one should specify
    \texttt{mag:\{max:20\}}. 
    To select only objects matching all criteria during 2 hours
    between the sunrise and sunset, the two parameters
    \texttt{duration:120, event:sun}
    must be specified.
  }

  \paragraph{Sorting options (\texttt{sort})}
  \add{The entries in the summary tables
    (see Table~\ref{tab:ex} for an example) are sorted in increasing
    right ascension by default.
    Changing this behavior is achieved by specifying 
    \texttt{sort=key}, in which \texttt{key} can take any value among 
    the reported parameters (Table~\ref{tab:col}).
    For instance, to sort object by increasing apparent magnitude, one
    should specify \texttt{sort=mv}.
    It is also possible to first group targets by type before applying
    the sorting, using \texttt{sort=type$>$key}.
  }

  \paragraph{Output formats (\texttt{mime})}
  \add{The results can be generated in three formats: VOTable, xHTML
    pages, and PDF file by specifying the
    \texttt{mime} parameter.
    Most users will likely use the PDF and xHTML outputs, for
    convenient display and interaction. The VOTable output has been
    designed for a more advanced usage, into a workflow for instance.
    It allows to directly query
    \vision~from, e.g., a software to control observing queues,
    retrieving the 
    values in the table, and the graphics for display within a
    graphical interface.
  }

  \paragraph{User identity (\texttt{from})}
  \add{We encourage users to provide their identity using the
    \texttt{from} parameter when they query \vision~via its Web
    service (option not available in the query form).
    We only keep this information for statistical purposes. It
    also allows to identify frequent users and their requests,
    in the case support is required.
  }

\section{Bridges to other services}
\label{sec:links}
  \indent To help further the observers in planning or conducting their
  observations, \vision~\add{dynamically builds links to three VO
  services for 
  each target, provided in the last three columns of the summary
  table
  (see the example in Table~\ref{tab:ex})}.
  \add{They provide, with a simple click, access to
    a view of the sky around the target in 
    Aladin (Sec.~\ref{sec:aladin}), 
    information on the target at CDS (Sec.~\ref{sec:cds}), 
    and detailed ephemerides at IMCCE (Sec.~\ref{sec:miriade}).
  }

\setlength{\titleOffset}{\hsize}
\settowidth{\titleLength}{Aladin at CDS\protect{\includegraphics[height=3ex]{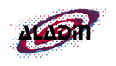}}}
\addtolength{\titleOffset}{-\titleLength}
\addtolength{\titleOffset}{-2em}

\begin{figure}[t]
    \centering
    \includegraphics[width=\hsize]{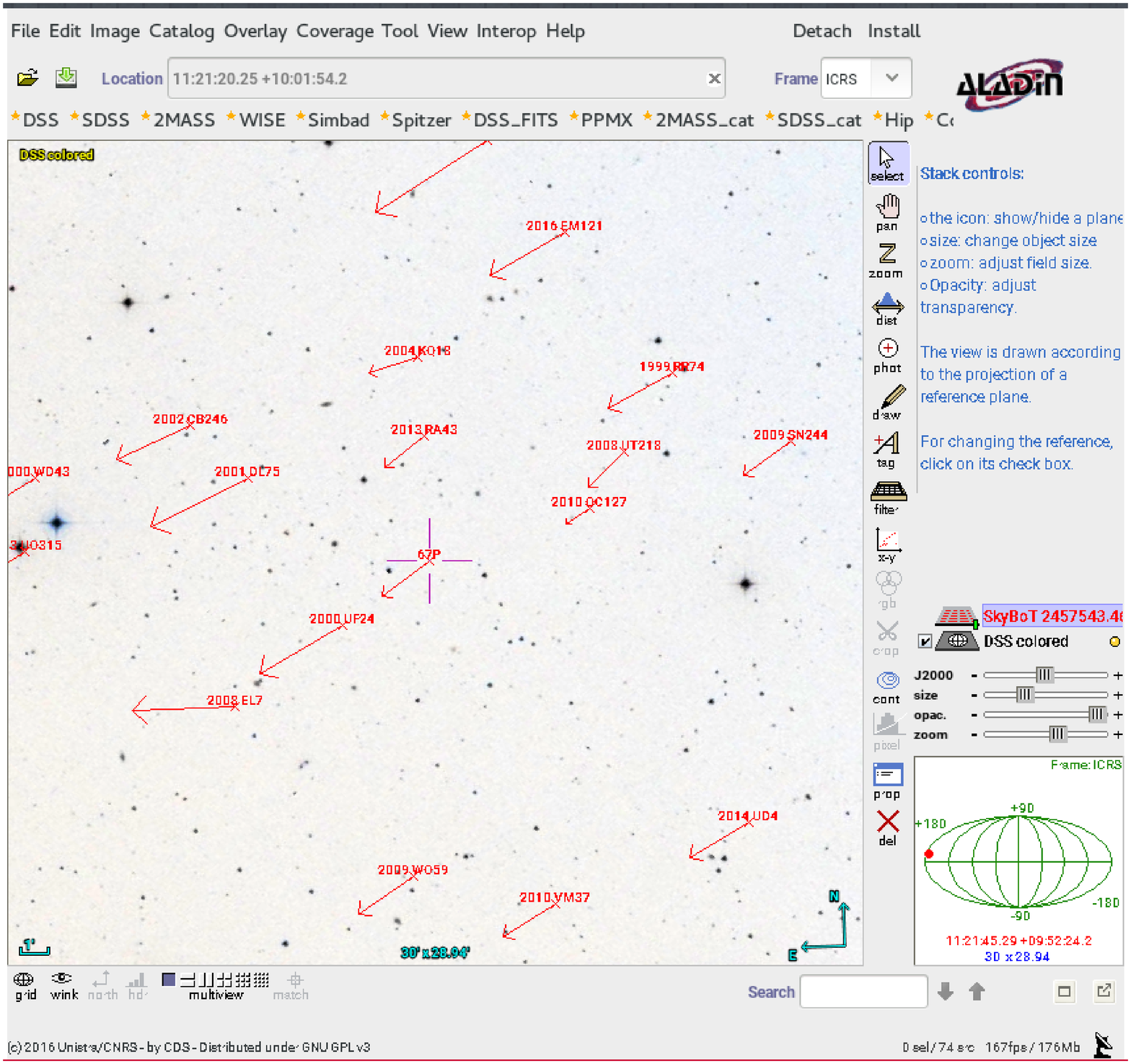}
    \caption{Example of the Aladin instance launched by \vision. Here
      the target was the comet 67P/Churyumov–Gerasimenko on June 4, 2016. 
      \label{fig:aladin}
    }
  \end{figure}

%  \subsection{Aladin at CDS\label{sec:aladin}} 
  \subsection[Aladin at CDS]%
             {Aladin at CDS\hspace{\titleOffset}\protect{\includegraphics[height=3ex]{logo-aladin}}\label{sec:aladin}} 
    \indent An interactive view of the sky around the target position
    is loaded in an
    instance of the Aladin Sky Atlas \citep{2000-AA-143-Bonnarel}, 
    host at the
    \textsl{Centre de Donn{\'e}es astronomiques de Strasbourg} (CDS).
    By default, \vision~\add{requests} the Digital Sky Survey 
    \add{(DSS)} to be
    displayed (Fig.~\ref{fig:aladin}),
    and overplots the known moving targets (at the time of
    passage at the maximum altitude) from the
    \skybot~service \citep{2006-ASPC-351-Berthier}.
    Users can then take advantage of all the tools offered by Aladin,
    such as overlaying additional catalogs and \add{fields of view} for instance,
    around their target position. 

%-------------------------------------------------------------
%-------------------------------------------------------------

\setlength{\titleOffset}{\hsize}
\settowidth{\titleLength}{Target information at CDS\protect{\includegraphics[height=3ex]{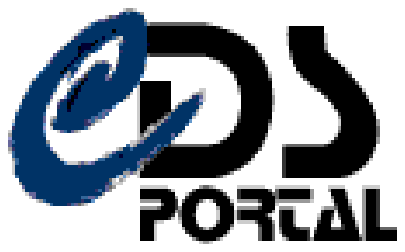}}}
\addtolength{\titleOffset}{-\titleLength}
\addtolength{\titleOffset}{-2em}

%  \subsection{Target information at CDS\label{sec:cds}} 
  \subsection[Target information at CDS]%
             {Target information at CDS\hspace{\titleOffset}\protect{\includegraphics[height=3ex]{logo-cds}}\label{sec:cds}} 
    \indent The orbital elements of the asteroids and comets
    from the \texttt{astorb} and \texttt{cometpro} catalogs
    hosted at CDS, used in the computation of ephemerides, are displayed 
    in a Web browser.
    For targets with fixed-coordinates, the \simbad~entry corresponding
    to their identifier is opened if resolved as a celestial object.

\setlength{\titleOffset}{\hsize}
\settowidth{\titleLength}{\miriade~at IMCCE\protect{\includegraphics[height=3ex]{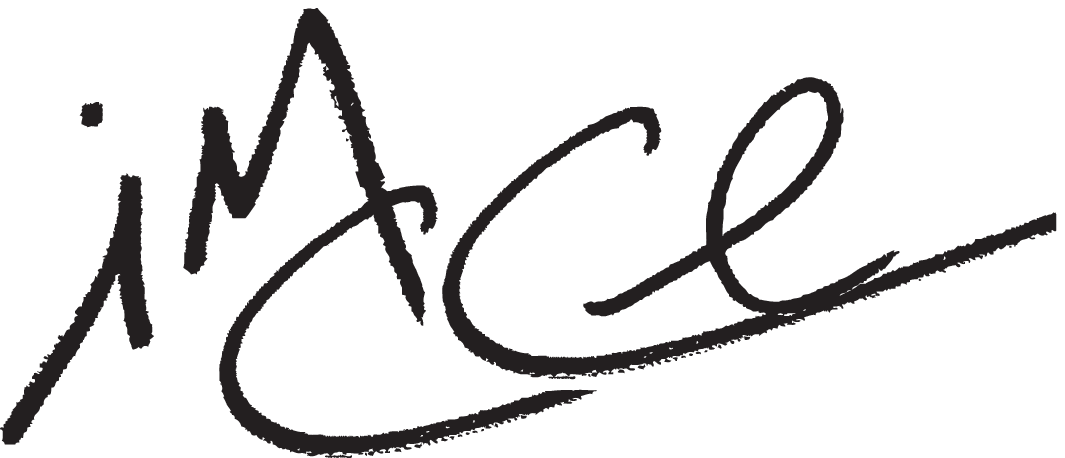}}}
\addtolength{\titleOffset}{-\titleLength}
\addtolength{\titleOffset}{-2em}
%  \subsection{\miriade~at IMCCE\label{sec:miriade}} 
  \subsection[\miriade~at IMCCE]%
             {\miriade~at IMCCE\hspace{\titleLength}\protect{\includegraphics[height=3ex]{logo-miriade}}\label{sec:miriade}} 
    \indent In the case of moving targets, the coordinates reported in
    the summary table at the time of highest altitude may not be
    sufficient to point the targets.
    This is the case for particularly fast-moving targets, such as
    near-Earth asteroids \add{(NEAs)},
    or for instruments with small fields of view.
    Ephemerides with a time step of
    10\,min can therefore be displayed in a Web browser, computed on
    the fly at 
    the IMCCE by the \miriade~service \citep[][]{2009-EPSC-Berthier}.
    \add{These} ephemerides are dedicated to observation, and local,
    horizontal, and equatorial coordinates are listed for the entire
    night.

\section{Conclusion\label{sec:conclu}}
  \indent We have created a freely accessible service, named \vision, 
  compliant with Virtual Observatory standards, which allows anyone to 
  create graphics and data tables of observing conditions of solar system 
  and (extra)galactic objects. 
  The access is standardized, and both command-line and Web
  form are available. Results can be returned in human-readable format
  (HTML, PDF) and VOTable for insertion in other programs, like
  telescope scheduler for instance.
  The results are tight with three other VO services, \add{hosted} at CDS and
  IMCCE to provide further information on targets and help in preparing
  the observations.

\section*{Acknowledgements}
  This development of this service was partly funded 
  by \add{Europlanet 2020 Research Infrastructure project of the
    European Union's Horizon 2020 research and 
    innovation programme under grant agreement No
    654208.  }

\bibliographystyle{elsart-harv} % style aa.bst
\bibliography{biblio}  % your references Yourfile.bib

  \begin{landscape}
\begin{table}
  \begin{center}
  \begin{tabular}{r@{\,}l|rr|r|rr|r@{$^h$\,}r@{$^m$\,}r@{$^s$\hspace{1em}}r@{\degr\,}r@{$^\prime$\,}r@{\arcsec\hspace{1em}}r|rr|rrrrr|ccc}
    \multicolumn{2}{c|}{Target}  &     m$_V$ & $\phi$\hspace{.75em} &     \multicolumn{1}{c|}{$\mathcal{D}$} &     Alt & Az. &     \multicolumn{3}{c}{RA}\hspace{2em} & \multicolumn{3}{c}{DEC}\hspace{1.5em} & Rate &     $\lambda_G$ & $\beta_G$ &     $r$\hspace{.75em} & $\Delta$\hspace{.75em} &     $\alpha$\hspace{.5em} & $\widehat{\rm SEO}$ & $\widehat{\rm MEO}$ &    \multicolumn{3}{c}{Links} \\ 
  \hline
 & Moon& -7.03&   1997&12$^h$30$^m$& 29& 61& 2&17&50& 10&08&50&    1772&335&-46& 0.002& 1.012&143.3& 36.6&  0.0&\href{http://vo.imcce.fr/webservices/miriade/ephemcc_query.php?-tscale=UTC&-mime=html&-tcoor=5&-step=10m&-nbd=145&-ep=2457541.37500000&-observer=309&-name=p:Moon}{\includegraphics[height=2ex]{logo-miriade}}&--&\href{http://aladin.u-strasbg.fr/java/nph-aladin.pl?-script=get%20hips(P/DSS2/color)%2034.459518%20+10.147369%2030.00000';get%20SkyBoT.IMCCE(2457541.951389,309,'Asteroids,%20Planets%20and%20Comets','120%20arcsec')%2034.459518%20+10.147369%2030.00000'}{\includegraphics[height=2ex]{logo-aladin}} \\
    67P & \truncate{3cm}{Churyumov-Gerasimenko}& 18.54&   0.11& 5$^h$49$^m$&55&359&11&20&43& 10&04& 6&      23&248& 62&  2.86&  3.12& 18.9& 95.3&139.8&\href{http://vo.imcce.fr/webservices/miriade/ephemcc_query.php?-tscale=UTC&-mime=html&-tcoor=5&-step=10m&-nbd=145&-ep=2457541.37500000&-observer=309&-name=c:67P}{\includegraphics[height=2ex]{logo-miriade}}&\href{http://vizier.u-strasbg.fr/viz-bin/VizieR-5?-ref=VIZ54f5b3529af2&amp;-source=B/comets/comets&Code=67P}{\includegraphics[height=2ex]{logo-cds}}&\href{http://aladin.u-strasbg.fr/java/nph-aladin.pl?-script=get%20hips(P/DSS2/color)%20170.179367%20+10.068503%2030.00000';get%20SkyBoT.IMCCE(2457541.472222,309,'Asteroids,%20Planets%20and%20Comets','120%20arcsec')%20170.179367%20+10.068503%2030.00000'}{\includegraphics[height=2ex]{logo-aladin}} \\
    & 12116& 18.09&   0.00& 7$^h$50$^m$&78&356&12&40&44&-13&37&24&       8&299& 48&  2.46&  3.12& 15.8&122.6&162.3&\href{http://vo.imcce.fr/webservices/miriade/ephemcc_query.php?-tscale=UTC&-mime=html&-tcoor=5&-step=10m&-nbd=145&-ep=2457541.37500000&-observer=309&-name=a:1999 JA34}{\includegraphics[height=2ex]{logo-miriade}}&\href{http://vizier.u-strasbg.fr/cgi-bin/VizieR-5?-source=B/astorb/astorb&Name=1999 JA34}{\includegraphics[height=2ex]{logo-cds}}&\href{http://aladin.u-strasbg.fr/java/nph-aladin.pl?-script=get%20hips(P/DSS2/color)%20190.183929%20-13.623556%2030.00000';get%20SkyBoT.IMCCE(2457541.527778,309,'Asteroids,%20Planets%20and%20Comets','120%20arcsec')%20190.183929%20-13.623556%2030.00000'}{\includegraphics[height=2ex]{logo-aladin}} \\
    & \truncate{3cm}{TYC5597-234-1}& 11.85&   0.00&10$^h$50$^m$&72&  2&15&46&30& -7&32&31&       0&360& 34&  1.00&  1.00&  0.0&160.2&153.4&--&\href{http://simbad.u-strasbg.fr/simbad/sim-basic?Ident=TYC5597-234-1}{\includegraphics[height=2ex]{logo-cds}}&\href{http://aladin.u-strasbg.fr/java/nph-aladin.pl?-script=get%20hips(P/DSS2/color)%20236.625366%20-7.542172%2030.00000';get%20SkyBoT.IMCCE(2457541.652778,309,'Asteroids,%20Planets%20and%20Comets','120%20arcsec')%20236.625366%20-7.542172%2030.00000'}{\includegraphics[height=2ex]{logo-aladin}} \\
    & \truncate{3cm}{Mars}& -1.97&  18.56&11$^h$00$^m$&86&348&15&40&47&-21&19&39&      53&347& 25&  0.50&  1.51&  9.4&165.9&152.5&\href{http://vo.imcce.fr/webservices/miriade/ephemcc_query.php?-tscale=UTC&-mime=html&-tcoor=5&-step=10m&-nbd=145&-ep=2457541.37500000&-observer=309&-name=p:Mars}{\includegraphics[height=2ex]{logo-miriade}}&--&\href{http://aladin.u-strasbg.fr/java/nph-aladin.pl?-script=get%20hips(P/DSS2/color)%20235.197220%20-21.327631%2030.00000';get%20SkyBoT.IMCCE(2457541.652778,309,'Asteroids,%20Planets%20and%20Comets','120%20arcsec')%20235.197220%20-21.327631%2030.00000'}{\includegraphics[height=2ex]{logo-aladin}} \\
    & \truncate{3cm}{M70}&  9.06&   0.00&10$^h$30$^m$&82&181&18&43&12&-32&17&31&       0&363&-13&  1.00&  1.00&  0.0&151.5&112.9&--&\href{http://simbad.u-strasbg.fr/simbad/sim-basic?Ident=M70}{\includegraphics[height=2ex]{logo-cds}}&\href{http://aladin.u-strasbg.fr/java/nph-aladin.pl?-script=get%20hips(P/DSS2/color)%20280.803192%20-32.292137%2030.00000';get%20SkyBoT.IMCCE(2457541.777778,309,'Asteroids,%20Planets%20and%20Comets','120%20arcsec')%20280.803192%20-32.292137%2030.00000'}{\includegraphics[height=2ex]{logo-aladin}} \\
    (107) & \truncate{3cm}{Camilla}& 13.07&   0.10& 8$^h$30$^m$&74&  2&19&57& 9& -9&05&38&      10&392&-19&  2.96&  3.71& 11.8&131.4& 95.0&\href{http://vo.imcce.fr/webservices/miriade/ephemcc_query.php?-tscale=UTC&-mime=html&-tcoor=5&-step=10m&-nbd=145&-ep=2457541.37500000&-observer=309&-name=a:Camilla}{\includegraphics[height=2ex]{logo-miriade}}&\href{http://vizier.u-strasbg.fr/cgi-bin/VizieR-5?-source=B/astorb/astorb&Name=Camilla}{\includegraphics[height=2ex]{logo-cds}}&\href{http://aladin.u-strasbg.fr/java/nph-aladin.pl?-script=get%20hips(P/DSS2/color)%20299.290283%20-9.094117%2030.00000';get%20SkyBoT.IMCCE(2457541.826389,309,'Asteroids,%20Planets%20and%20Comets','120%20arcsec')%20299.290283%20-9.094117%2030.00000'}{\includegraphics[height=2ex]{logo-aladin}} \\
    & \truncate{3cm}{Triton}& 13.52&   0.12& 5$^h$40$^m$&73&359&22&54& 2& -7&54&59&       0&243&-56& 29.93& 29.96&  1.9& 90.3& 53.8&\href{http://vo.imcce.fr/webservices/miriade/ephemcc_query.php?-tscale=UTC&-mime=html&-tcoor=5&-step=10m&-nbd=145&-ep=2457541.37500000&-observer=309&-name=p:Triton}{\includegraphics[height=2ex]{logo-miriade}}&--&\href{http://aladin.u-strasbg.fr/java/nph-aladin.pl?-script=get%20hips(P/DSS2/color)%20343.511993%20-7.916412%2030.00000';get%20SkyBoT.IMCCE(2457541.951389,309,'Asteroids,%20Planets%20and%20Comets','120%20arcsec')%20343.511993%20-7.916412%2030.00000'}{\includegraphics[height=2ex]{logo-aladin}} \\
    & SMC&  0.00&   0.00& 7$^h$30$^m$&39&168& 0&53&59&-72&00& 0&       0&302&-45&  1.00&  1.00&  0.0&101.9& 83.3&--&--&\href{http://aladin.u-strasbg.fr/java/nph-aladin.pl?-script=get%20hips(P/DSS2/color)%2013.500000%20-72.000000%2030.00000';get%20SkyBoT.IMCCE(2457541.951389,309,'Asteroids,%20Planets%20and%20Comets','120%20arcsec')%2013.500000%20-72.000000%2030.00000'}{\includegraphics[height=2ex]{logo-aladin}} \\
  \hline
  \end{tabular}
  \caption{Example of the summary table associated with
  Figs.~\ref{fig:airmass} and~\ref{fig:sky}.
   Values are reported at the time of the highest altitude. See Table
  1 for a description of each column.
  \label{tab:ex}}
  \end{center}
\end{table}
\end{landscape}

\end{document}